\documentstyle[12pt,twoside,fleqn,espcrc1,epsfig]{article}

\newcommand{\beq}{\begin{equation}}
\newcommand{\eeq}{\end{equation}}
\newcommand{\bea}{\begin{eqnarray}}
\newcommand{\eea}{\end{eqnarray}}

\title{Chiral Symmetry Restoration in the Instanton Liquid 
       at Finite Density}

\author{Ralf Rapp   
        \thanks{supported by the A.-v.-Humboldt foundation as a Feodor-Lynen
                fellow and by the US-Department of Energy under grant no.
                DE-FG02-88ER40388.} \\ 
Department of Physics and Astronomy,
        State University of New York at Stony Brook, \\
        Stony Brook, NY 11794-3800, U.S.A.}

\begin{document}
\maketitle

\begin{abstract}
The properties of the QCD partition function at finite chemical 
potential are studied within the instanton liquid model. It is shown 
that the density dependence of the quark-induced instanton-antiinstanton
($I$-$A$) interaction leads to the formation of topologically neutral 
$I$-$A$ pairs ('molecules'), 
resulting in a first order chiral phase transition at a critical chemical 
potential $\mu_q^c\simeq 310$~MeV. At somewhat higher densities
($\mu_q\ge360$~MeV), the quark Fermi surface becomes instable with respect 
to diquark condensation (Cooper pairs) generating  BCS-type energy gaps
of order 50~MeV.   
\end{abstract}

\section{INTRODUCTION}
The investigation of the phase diagram of QCD is one of the 
central issues in understanding the properties of strong
interactions. While the finite temperature axis has been 
theoretically explored in quite some detail, much less
is known about the finite density ($\mu_q$-) axis,   
which is partly due to the fact that first principle QCD lattice
calculations at finite $\mu_q$ encounter the problem of a complex
fermionic determinant when integrating the QCD partition function. 
The phase structure of QCD at finite density, however, might be very 
rich, including new forms of condensates (other than the ordinary 
chiral condensate present in the QCD vacuum), different orders of 
phase transitions in the $\mu_q$-$T$-plane (entailing a tricritical 
point~\cite{HaSt}), etc..      
In this contribution, however, we will focus on the T=0-, $\mu_q$$>$0-
axis and try to examine the nature of the chiral phase transition 
in this regime. We will do so within the framework of the instanton
liquid model (ILM) of QCD, which, although lacking explicit confinement, 
yields a very successful phenomenology of the QCD vacuum structure, 
and the low-lying hadron spectrum. Recently it also provided an interesting
mechanism for chiral symmetry restoration at finite temperature, based
on the rearrangement of the (anti-) instantons within the liquid, 
rather than on a mere disappearance of them as suggested earlier.   
Our objective here is to investigate whether a similar mechanism
could be responsible for the restoration of chiral symmetry
at finite density as well.  

We start by briefly recalling some features of the instanton liquid model 
at zero density (sect.~2) and then turn to the finite 
density case (sect.~3), where we
first calculate the $\mu_q$-dependence of the instanton-antiinstanton 
interaction and then assess its impact on the chiral phase transition. 
Under certain approximations
we will be able to avoid a complex partition function and estimate the
critical chemical potential within a mean-field type approach. 
We furthermore include  some correlations between quarks at the
Fermi surface and show how the recently discussed mechanism of  
color superconductivity~\cite{ARW,RSSV} figures in our 
mean-field description.

\section{THE INSTANTON LIQUID MODEL AT ZERO DENSITY}

The instanton model in QCD is based on the assumption that the 
vacuum is dominated by non-peturbative gauge field configurations
which constitute semiclassical ('instanton') solutions of the 
Yang-Mills equations. 
The QCD partition function in the instanton ap\-prox\-imation then becomes
\beq
Z_{QCD}^{inst}=\sum\limits_{N_+,N_-} \frac{1}{N_+! N_-!} 
\prod\limits_{I=1}^{N_+,N_-} \int d\Omega_I \ n(\rho_I) \ 
e^{-S_{int}} \ \rho_I^{N_f} \prod\limits_{f=1}^{N_f} 
det (i \not\!\!D+im_f)  \ ,  
\eeq 
where the path-integral over all possible field configurations has been  
converted into an integration over the so-called collective coordinates
$\Omega_I=\{z_I,\rho_I,u_I \}$ (position, size and color orientation) 
of $N_+$ instantons and $N_-$ 
antiinstantons. The single-instanton amplitude $n(\rho_I)$ (including 
quantum corrections) and the gluonic part of the instanton interaction
$S_{int}$ determine the total instanton density in the 
absence of ferminons to be $N/V=2\int~d\rho~n(\rho)~e^{-S_{int}}$. 
The determinant of the Dirac operator, arising from the integration
over the quark fields, is approximately calculated by keeping only the  
lowest lying modes, which should be reasonbale for assessing 
the bulk properties of the system. The Dirac equation in 
the (anti-) instanton field possesses 
a (right-) left-handed zero energy solution. In the basis spanned by 
these zero modes, $\Psi_{0,I}$ and $\Psi_{0,A}$, and neglecting small 
current quark masses ($m_f\to 0$), the fermionic determinant reads  
\bea
det(i\not\!\!D) \simeq det \left( \begin{array}{cc} 0 & T_{IA} \\ 
                                                T_{AI} & 0 
                              \end{array} \right) 
         =|T_{IA}|^2 \ ,  
\eea 
where the overlap matrix element 
\beq
T_{IA}(z,u)=\int d^4x \ \Psi_{0,I}^\dagger(x-z_I,u_I) \
(i\not\!\!{D}) \ \Psi_{0,A}^\dagger(x-z_A,u_A)
\equiv i \ u\cdot \hat{z} \ f(z) \ 
\label{Tia0} 
\eeq
is linear in the relative $SU(3)$-color orientation characterized by a 
complex four vector
$u_\mu$ ($z=z_A-z_I$ denotes the relative distance between $I$ and  
$A$). Lorentz invariance then implies that $T_{IA}$ is 
determined by a single scalar function $f(\sqrt{z^2})$. 

With the key parameters taken as $N/V\simeq$~(1--1.4)~fm$^{-4}$ and 
$\rho_I=\rho_A\simeq$~1/3~fm (which have also been confirmed in
lattice calculations) a successful phenomenology of the QCD vacuum
and the low-lying hadronic spectrum can be obtained~\cite{SS98}.

\section{THE INSTANTON LIQUID MODEL AT NON-ZERO DENSITY}

\subsection{$I$-$A$ Interaction at Finite $\mu_q$: 
Quark Zero Modes and $T_{IA}$}
To study medium modifications of the instanton ensemble 
we first have to construct their interactions at finite quark 
chemical potential. We start from the finite density Dirac equation 
(in euclidean space), which still has zero mode solutions satisfying
\beq
(i\not\!\!D_I-i\mu_q\gamma_4)\Psi_{0,I}=0 \ .
\eeq
Constraining ourselves to zero temperature, 
the gluonic (instanton-) fields entering the covariant derivative 
are not affected by $\mu_q$. The explicit form of the quark zero modes 
has been determined in ref.~\cite{Ab83},  
\beq
\Psi_{0,I}(|\vec x|,x_4;\mu_q)=\frac{1+\gamma_5}{2} \ \frac{\rho}{2\pi} \ 
\frac{{\rm e}^{\mu_q x_4}}{\Pi^{1\over 2}(x)} \ \left(\not\!\partial-
\frac{\not\!\partial \Pi(x)}{\Pi(x)}\right) \
\frac{\cos(\mu_q |\vec x|)+\frac{x_4}{r} \sin(\mu_q |\vec x|)}{x^2} \ 
{\rm e}^{-\mu_q x_4} \ \chi
\eeq  
with the function $\Pi(x)=(1+\rho^2/x^2)$ and a spin-color spinor
$\chi$. Note that the solution of the adjoint Dirac equation, 
\beq
\Psi_{0,I}^\dagger(x;-\mu_q) \ (i\not\!\!D_I-i\mu_q\gamma_4)=0 \ ,
\eeq 
carries the chemical potential argument with opposite sign. This 
is necessary for a consistent definition of expectation values 
at finite $\mu_q$, and in particular renders a finite norm 
\beq
\int d^4x \ \Psi_{0,I}^\dagger(x;-\mu_q) \ \Psi_{0,I}(x;\mu_q) =1 \ . 
\eeq

With the properly constructed quark wave functions we can now evaluate 
the fermionic overlap matrix element representing the zero mode part
of the full Dirac operator. Choosing for simplicity the sum ansatz
for the gauge-field configurations, $A_\mu=A_\mu^I+A_\mu^A$, entering 
the covariant derivative, allows us to replace the latter by an ordinary
one, yielding 
\bea
T_{IA}(z,u;\mu_q) & = & - \int d^4x \ 
\Psi_{0,I}^\dagger(x-z_I;-\mu_q) \ (i\not\!\partial-i\mu_q\gamma_4) \ 
\Psi_{0,A}(x-z_A;\mu_q) 
\nonumber\\ 
 & \equiv & i \ u_4 \ f_1(\tau,r;\mu_q) + i \ 
\frac{(\vec u \cdot \vec r)}{r} \ f_2(\tau,r;\mu_q) \ .  
\label{Tiamu}
\eea
The breaking of Lorentz 
invariance in the medium implies the existence of two independent 
(real-valued) functions $f_1, f_2$ which are shown in fig.~1.   
\begin{figure}[htb]
{\makebox{\epsfig{file=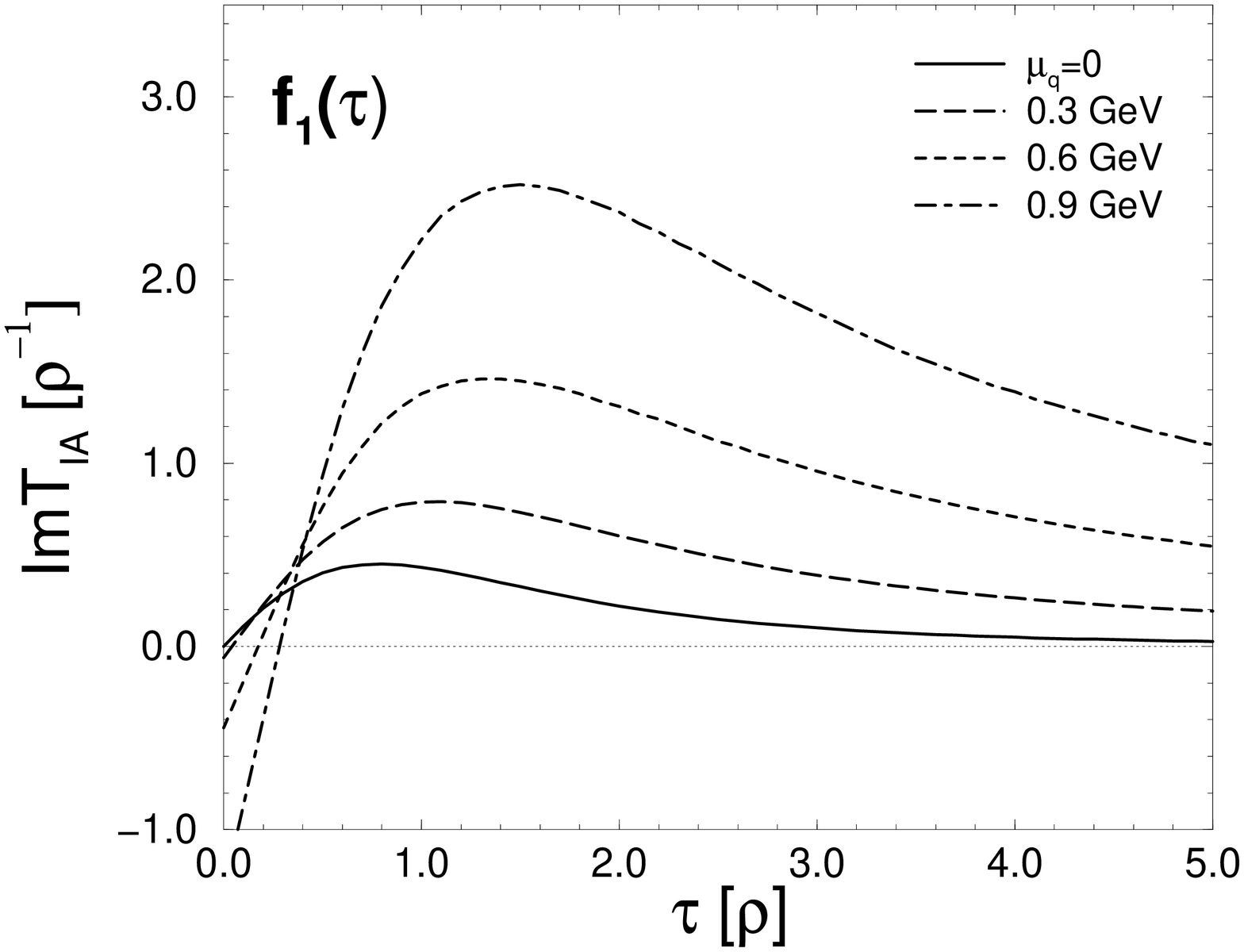,width=65mm,angle=-90}}}
\hspace{0.15cm} 
{\makebox{\epsfig{file=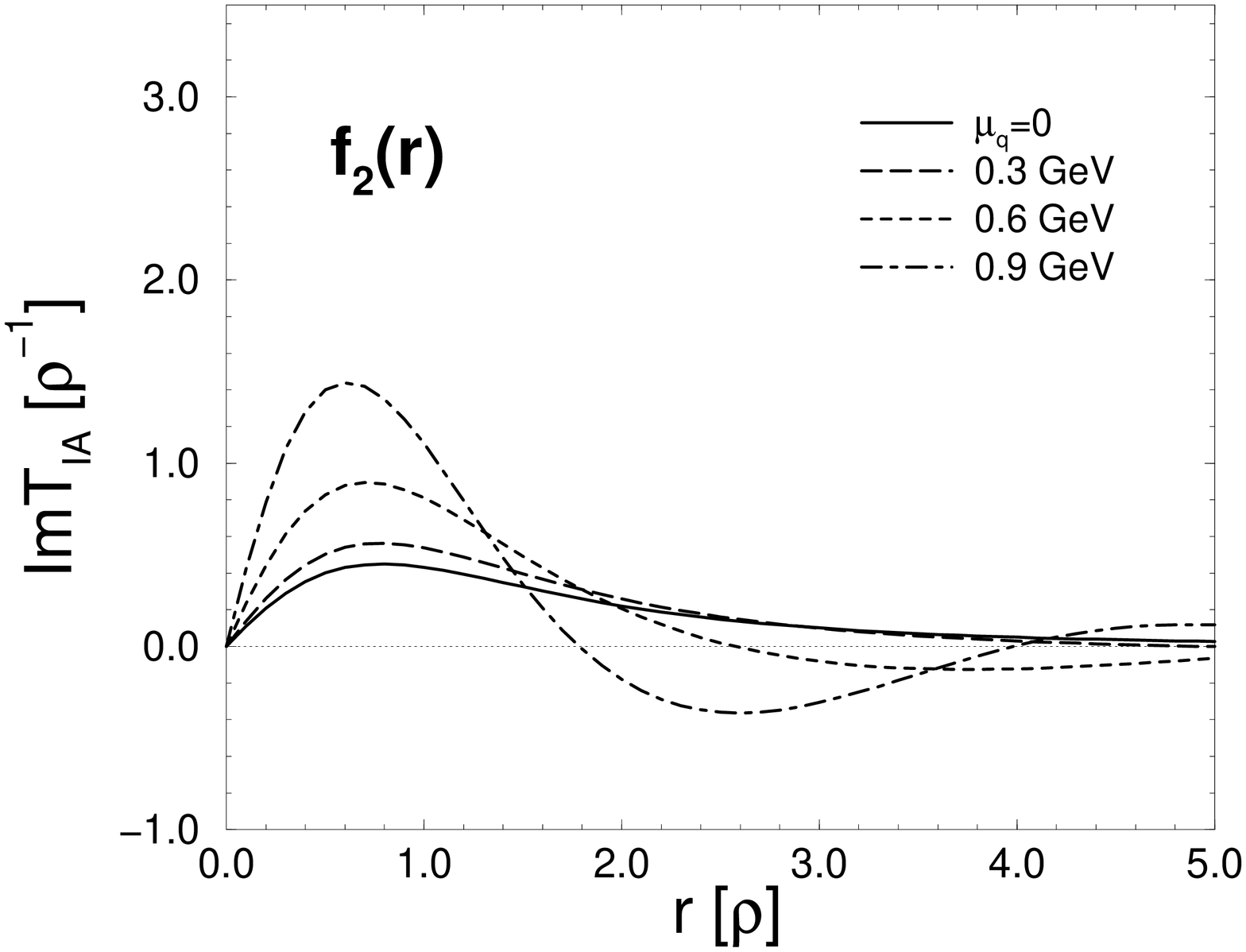,width=65mm,angle=-90}}}
\caption{Quark-induced $I$-$A$-interaction at finite density 
for the most attractive color orientation $u_4$=1, $\vec u$=0 as well 
as $r$=0 (left panel) and for $u_4$=0, $|\vec u|$=1 and $\tau$=0 
(right panel). }
\end{figure}
Similar to what has been found at finite temperature~\cite{SV91}, $T_{IA}$
is strongly enhanced 
in  temporal direction with increasing $\mu_q$, but  
oscillates as $\sim \sin(2\mu_q r)$ in spatial direction~\cite{S97}. 
The latter effectively suppresses the interaction once integrating 
over $r$. 


\subsection{Thermodynamics of the ILM at Finite $\mu_q$}
To investigate the finite density properties of the thermodynamic
potential (or free energy)  we here resort to the
so-called cocktail model introduced in ref.~\cite{IS94}. 
It amounts to a mean-field type description including three major 
components in the system: essentially random (anti-) instantons
(the 'atomic' component), strongly correlated $I$-$A$-pairs (the
'molecular' component) and a Fermi sphere of constituent quarks
('quasiparticles'). Thus
\beq
\Omega(\mu_q)=\Omega_{inst}(\mu_q) + \Omega_{quark}^{QP}(\mu_q) \ ,   
\label{Omega} 
\eeq
where the constituent quark contribution  
\beq
\Omega_{quark}^{QP}(m_q;\mu_q)  =  \epsilon_q(m_q;\mu_q) 
-\mu_q \ n_q(m_q;\mu_q)
\eeq
is zero for Fermi energies $\mu_q<m_q$, with
$m_q$ denoting the constituent quark mass. The instanton part
of the free energy,  
\beq
\Omega_{inst}(n_a,n_m;\mu_q)=-\frac{\ln[Z_{inst}(n_a,n_m;\mu_q))]}{V_4}
=-n_a \ \ln\left[ \frac{e z_a}{n_a} \right] 
- n_m \ \ln\left[ \frac{e z_m}{n_m} \right] \  , 
\eeq
is related to the atomic and molecular 'activities'
\bea
z_a & = & 2 \ C \ \rho^{b-4} \ {\rm e}^{-S_{int}} \ 
\langle T_{IA}(\mu_q) T_{AI}(\mu_q)\rangle^{N_f/2} 
\nonumber\\
z_m & = & C^2 \ \rho^{2(b-4)} \ {\rm e}^{-2S_{int}} \
\langle [T_{IA}(\mu_q) T_{AI}(\mu_q)]^{2N_f}\rangle \ .  
\label{z}
\eea
The minimization of $\Omega$ with respect to the corresponding 
(4-dimensional) densities $n_a$ and $n_m$ determines the equilibrium 
state of the system at fixed $\mu_q$. In particular, 
$\Omega_{inst}$ encodes all the features 
of the $T=\mu_q=0$ instanton ensemble, as {\it e.g.} the quark condensate
and the constituent quark mass, which in mean-field approximation 
are given by $\langle \bar qq\rangle =-1/(\pi\rho) (3/2 n_a)^{1/2}$ 
and $m_q\propto -\rho^2 \langle \bar qq\rangle $. Therefore the
normalization constant $C\propto (\Lambda_{QCD})^b$ can be fixed
to give $N/V$=1.4~fm$^{-4}$, being realized for $n_a$=1.34~fm$^{-4}$
and $n_m$=0.03~fm$^{-4}$, which is not unreasonable.    
The gluonic interaction has been approximated by an average
repulsion $S_{int}=-\kappa \rho^4(n_a+2n_m)$. \\ 
At finite $\mu_q$ the Dirac operator is not hermitian
any more, {\it i.e.} $T_{AI}(\mu_q)\ne T_{IA}^\dagger(\mu_q)$, 
resulting, in general, in a complex 
fermionic determinant, entailing the well-known 'sign' problem. However, 
assuming an average gluonic interaction that
does not depend on density allows us to perform the color averages
implied in eqs.~(\ref{z}) analytically, {\it i.e.} 
\bea
z_a & \propto & \int du T_{IA}(\mu_q) T_{IA}^\dagger(-\mu_q) 
= \frac{1}{2N_c} [f_1^+f_1^-+f_2^+f_2^-] 
\nonumber\\
z_m & \propto & \int du [T_{IA}(\mu_q) T_{IA}^\dagger(-\mu_q)]^{N_f}
=\frac{ \left[ (2N_c-1) \{f_1^+f_1^-+f_2^+f_2^-\}^2
+\{f_1^+f_2^-f_1^-f_2^+\}^2\right]}{4N_c(N_c^2-1)}
  \   
\eea 
($N_f=2$, $f_i^{\pm}\equiv f_i(\pm\mu_q)$),  
which ensures the pressure to remain real. Fig.~2 shows our results 
as function of the quark chemical potential (left panel) for 
$N_c=3, N_f=2$.
At small $\mu_q$ essentially nothing happens until, at a critical
value $\mu_q^c\simeq 310$~MeV, the system jumps into the chirally
restored phase, the latter being  characterized by $n_a=0$. 
The transition is of first order, as can be seen by inspection of the 
$m_q$-dependence of the free energy (right panel of fig.~2).
\begin{figure}[!t]
{\makebox{\epsfig{file=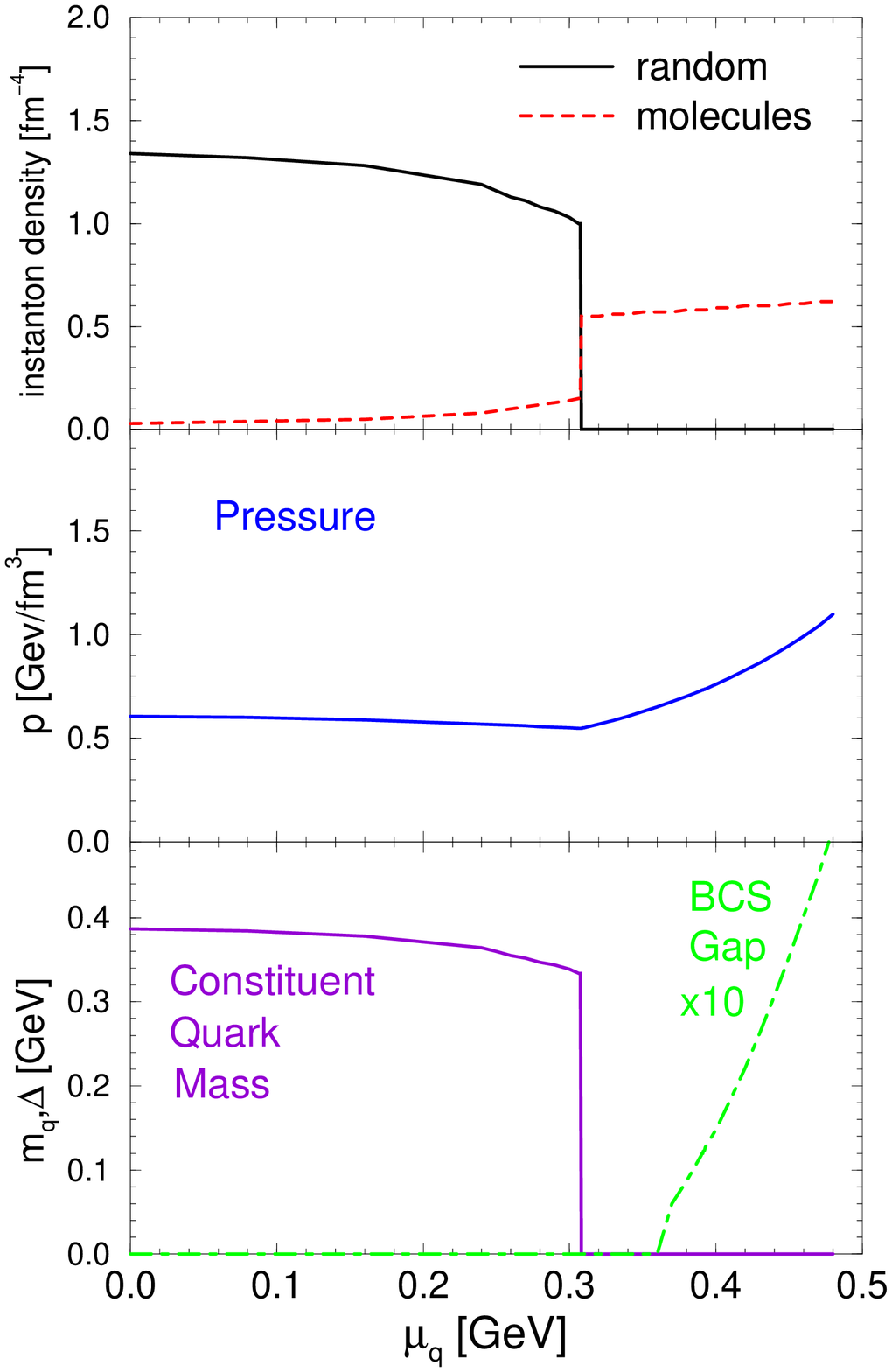,width=90mm,angle=0}}}
\hspace{-0.7cm}
{\makebox{\epsfig{file=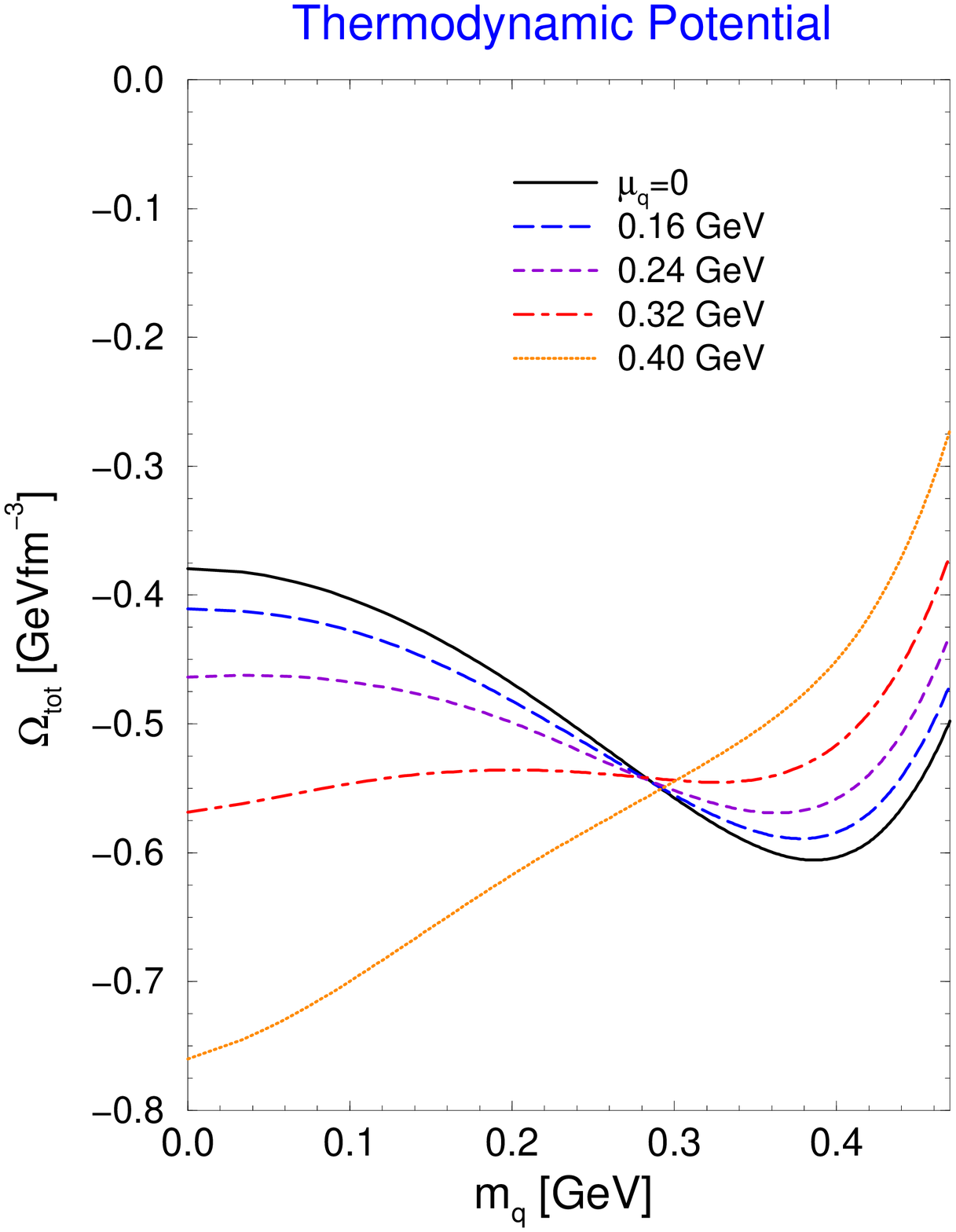,width=75mm,angle=0}}}
\caption{Our results for the instanton ('atomic')
and molecule densities (upper left panel), 
the pressure $p=-\Omega$ (middle left panel) and the constituent quark mass
(lower left panel) after minimizing the free energy, eq.~(\ref{Omega}),   
with respect to $n_a$ and $n_m$; additionally including scalar diquark
correlations (as discussed in sect.~3.3) adds the dashed-dotted line to 
the lower left panel, representing the BCS energy gap at the quark Fermi 
surface.  The right panel shows the free energy  
as a function of constituent quark mass $m_q \propto n_a^{1/2}$,  
indicating a first order transition from 
the minimum at finite $m_q$ to the one at $m_q=0$. }
\end{figure}
Below $\mu_q^c$, the pressure actually decreases slightly with increasing 
$\mu_q$ indicating a mixed phase-type instability, similar to what 
has been discussed in refs.~\cite{ARW,Buba}. A significant
difference, however, is given by the fact that in our approach 
the total instanton density at the transition (residing in 
$I$-$A$-molecules) is still appreciable, $N/V=2n_m\simeq 1.1$~fm$^{-4}$, 
providing the major part of the pressure at this point; in other words: 
a substantial part of the nonperturbative vacuum pressure persists 
in the chirally restored phase (of course, eventually it will be  
suppressed due to the Debye screening of the instanton fields, which
we have not included here).

\subsection{Color Superconductivity}
As has recently been pointed out in refs.~\cite{ARW,RSSV}, the 
quark Fermi surface in the plasma phase might be unstable with 
respect to the formation of quark-Cooper pairs, once an attractive
$q$-$q$ interaction is present. Using the instanton-induced interaction 
in the scalar, color-antitriplett diquark channel (which is essentially   
the Fierzed-transformed interaction leading to a deeply bound pion and
is also phenomenolgically well-supported by baryonic spectroscopy),  
BCS-type energy gaps of $\Delta_0\simeq$~50-100~MeV have been 
predicted.\\ 
In the mean-field approach employed here, this interaction modifies 
the quark quasiparticle contribution according to
\beq
\Omega_q^{\Delta}(\mu_q)=tr \log\left[D(m_q,\Delta_0)\right] -
tr \left[D(m_q,\Delta_0) \Sigma(\Delta_0)\right] +
G tr\left[F(m_q,\Delta_0)\right] tr\left[\bar F(m_q,\Delta_0)\right] 
\label{Omqdel} 
\eeq
with the quasiparticle quark-propagator $D$ and the annomalous (Gorkov)
propagator $F$ corresponding to creation/annihilation of a bound 
Cooper pair. $G$ is the effective coupling constant derived from 
the instanton $q$-$q$-vertex~\cite{RSSV}. 
From $\Omega_q^\Delta$ one can obtain the standard gap equation for 
$\Delta_0$. However, here we also have to account for the fact that
a finite $\Delta_0$ will damp the quark zero-mode propagators in the
instanton part of the free energy, which results in a suppression 
of the quark induced $I$-$A$ interaction $T_{IA}$ of eq.~(\ref{Tiamu}). 
Thus, $\Omega_{inst}$ {\it disfavors} finite $\Delta_0$'s. 
The resulting expression for the free energy is of the form
\beq
\Omega(\mu_q)= \Omega_{inst}(n_a,n_m,\Delta_0;\mu_q) 
+\frac{1}{N_c} \left[2\Omega_q^\Delta(m_q,\Delta_0;\mu_q)+(N_c-2)
\Omega_q(m_q;\mu_q)\right] , \  
\eeq 
(the last term accounting for unpaired quarks), which now has to be 
minimized {\it w.r.t.}~$n_a$, $n_m$ and $\Delta_0$. 
We find that color superconductivity does not appear before 
$\mu_q\simeq$~360~MeV (see dashed-dotted curve in the lower left panel 
of fig.~2): although the quark-part of the free energy by itself, 
eq.~(\ref{Omqdel}), always favors a finite value for $\Delta_0$, the 
suppression caused by $\Delta_0$ in $\Omega_{inst}$ ({\it i.e.} the 
damping of quark-propagation in $T_{IA}$) prevents the formation 
of a diquark condensate below $\mu_q\simeq$~360~MeV. This again 
reflects the fact that the instanton contribution to the pressure is 
still dominant in the region somewhat above $\mu_q^c$, so that the  
gain due to a finite $\Delta_0$ in $\Omega_q^\Delta$ cannot overcome
the 'penalty' in $\Omega_{inst}$. Above $\mu_q\simeq$~360~MeV energy
gaps of order $\sim$~50~MeV arise, in line with the findings of 
ref.~\cite{RSSV}.\\
As far as the  thermodynamic properties of the system are concerned, 
no changes as compared to the previous section occur below 
$\mu_q=~360~MeV$, and even above the results for $n_a(\mu_q)$, 
$n_m(\mu_q)$, $p(\mu_q)$ and $m_q(\mu_q)$ would be hardly 
distinguishable from the curves displayed in fig.~2.  

 
\vspace{0.5cm} 
In summary, we have studied the QCD partition function at finite 
density using the instanton model.
Employing a simple mean-field approach, chiral symmetry restoration 
emerges at $\mu_q^c$~=310~MeV as a first order transition from an 
essentially random (anti-/) instanton liquid into a phase of strongly 
correlated $I$-$A$ molecules (and massless quarks), driven by the 
density dependent increase of the 
quark-induced $I$-$A$-interaction. Neglecting possible medium modifications
in the gluonic interaction, which are expected to be small, 
 a complex pressure could be avoided. 
Accounting for correlations in the scalar diquark channel, color
superconductivity sets in for chemical potentials $\mu_q\ge$~360~MeV,
associated with BCS gaps of up to $\sim$~50~MeV.   
 
\vspace{0.5cm} 

{\bf Acknowledgement} \\ 
I thank my collaborators E. Shuryak, T. Sch\"afer and M. Velkovsky
for fruitful discussions.

\end{document}